\documentclass[twocolumn,aps,pra,showpacs,floatfix,superscriptaddress]{revtex4-1}
\usepackage{amssymb,amsthm, amssymb, latexsym}
\usepackage{graphicx}
\usepackage{amsmath}
\usepackage[english]{babel}
\usepackage{color}

\newcommand{\ket}[1]{\left| #1 \right\rangle}

\begin{document}
\title{Half-integer Mott-insulator phases in the imbalanced honeycomb lattice}

\author{Krzysztof Gawryluk}
\affiliation{Centre for Quantum Technologies, National University of Singapore, 3 Science Drive 2, Singapore 117543, Singapore}
\affiliation{Wydzia{\l} Fizyki, Uniwersytet w Bia{\l}ymstoku, ul. Lipowa 41, 15-424 Bia{\l}ystok, Poland}

\author{Christian Miniatura}
\affiliation{Institut Non Lin\'{e}aire de Nice, UMR 7335, UNS, CNRS; 1361 route des Lucioles, 06560 Valbonne, France}
\affiliation{Merlion MajuLab, CNRS-UNS-NUS-NTU International Joint Research Unit UMI 3654, Singapore}
\affiliation{Centre for Quantum Technologies, National University of Singapore, 3 Science Drive 2, Singapore 117543, Singapore}
\affiliation{Department of Physics, National University of Singapore, 2 Science Drive 3, Singapore 117542, Singapore}

\author{Beno\^{\i}t~Gr\'{e}maud}
\affiliation{Laboratoire Kastler Brossel, Ecole Normale Sup\'{e}rieure CNRS, UPMC; 4 Place Jussieu, 75005 Paris, France}
\affiliation{Merlion MajuLab, CNRS-UNS-NUS-NTU International Joint Research Unit UMI 3654, Singapore}
\affiliation{Centre for Quantum Technologies, National University of Singapore, 3 Science Drive 2, Singapore 117543, Singapore}
\affiliation{Department of Physics, National University of Singapore, 2 Science Drive 3, Singapore 117542, Singapore}

\date{\today}

\begin{abstract}
Using mean-field theory, we investigate the ground state properties of ultracold bosons loaded in a honeycomb lattice 
with on-site repulsive interactions and imbalanced nearest-neighbor hopping amplitudes. 
Taking into account correlations between strongly coupled neighboring sites through an improved Gutzwiller ansatz, 
we predict the existence of half-integer Mott-insulator phases, i.e. states with half-integer filling and vanishing 
compressibility. These insulating phases result from the interplay between quantum correlations and the topology of 
the honeycomb lattice, and could be easily addressed experimentally as they have clear signatures in momentum space.
\end{abstract}


\maketitle

\section{Introduction}

Because of its remarkable low-energy electronic excitations, graphene has been the source of many key discoveries~\cite{Novoselov_04,Neto_09} 
which have sparked a vivid research flow now reaching new territories, as exemplified by ultracold atoms loaded in 
optical lattices~\cite{Zhu_07,grapheneKL, muramatsu,Sengstock_11_a,Sengstock_11_b, Esslinger1,Esslinger2,Esslinger3}. In this paper, we address 
the bosonic Mott-insulator to superfluid (MI-SF) transition taking place in the honeycomb lattice \cite{Kimchi,Chen11} 
and show that the phase diagram is richer than for the square lattice \cite{Jaksch98,Sachdev,expBloch,expPhillips,expPhillips2,bosonsGeorge}. 
Indeed, being genuinely bipartite, the honeycomb lattice has a 2-site (labeled \textsc{a} and \textsc{b}) unit Bravais 
cell which can accommodate symmetric and antisymmetric states. 
This has dramatic consequences for the ground state  of the interacting system, in either the Mott or superfluid phases;
Strikingly, \textit{half-integer} 
Mott lobes develop when nearest-neighbor hopping amplitudes are imbalanced.
This situation is similar to the Kagome lattice~\cite{kagome,kagome2,kagome3}, or more generally to any lattice for which the unit cell
comprises more than one site, a situation which is unavoidable in the presence of an external magnetic field.

The paper is organized as follows. In section~\ref{model}, we introduce the model and the extended Gutzwiller 
method~\cite{tcr,ageorge_varenna,tutorialLew} needed to correctly capture the
inter-site correlations responsible for these new half-integer Mott lobes.
In section~\ref{udim}, we discuss the uncoupled dimer solutions, i.e. the properties of the Mott phases, at integer and half-integer fillings. 
In section~\ref{cdim}, we present our numerical results for the coupled dimers, in particular the transition from the Mott phase  to
the superfluid phase. In section~\ref{crithop}, we discuss in more details the boundary of the half-filling Mott lobe properties, which can be obtained
analytically emphasizing the transition from a quasi-1D situation to a 2D square lattice phase diagram. 
The experimental signatures in momentum space, i.e. in the velocity distribution, of the different phases  are discussed in  section~\ref{sec:expsign}.
A summary of results and conclusions
are given in section~\ref{conc}.

\section{Model and methods}
\label{model}

\begin{figure}[thb]
\includegraphics[width=6cm]{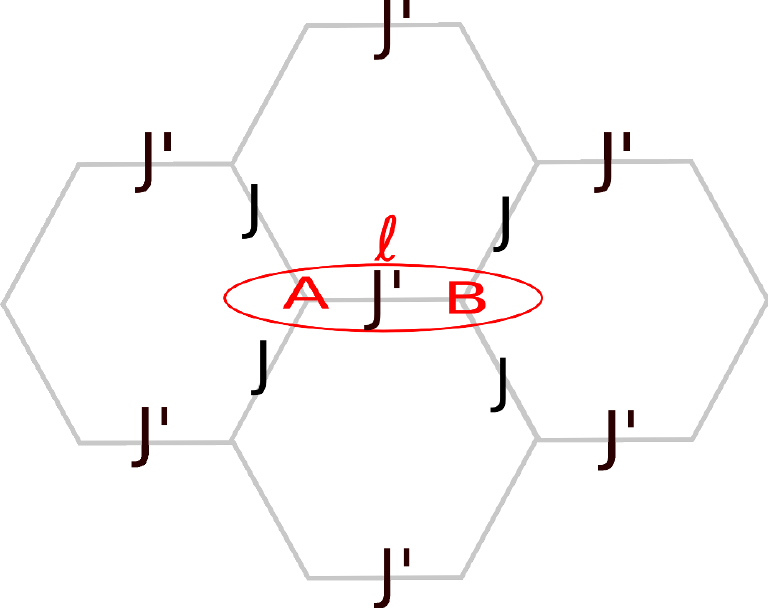}
\caption{\label{graph_lattice}(color online). Honeycomb lattice geometry with imbalance hopping amplitudes $J'$ and $J$. 
When $J'\gg J$, the system can be described by weakly coupled $\textsc{a}-\textsc{b}$ dimers living on the horizontal links denoted by $\ell$.   }
\end{figure}

Let us consider interacting bosons loaded on the honeycomb lattice with nearest-neighbor tunneling and further assume 
that one hopping parameter $J'$ is different from the two other (identical) ones $J$, see Fig~\ref{graph_lattice}.

The limit $J' \ll J$ corresponds to weakly-coupled 1D-chains, whereas the limit $J' \gg J$ corresponds to weakly-coupled dimers. 
It is worth noticing that hopping imbalance in the graphene lattice has already been achieved 
experimentally~\cite{Esslinger1,Esslinger2,Esslinger3,Sengstock_11_a,Sengstock_11_b}. 
In the following we single out neighboring \textsc{a} and \textsc{b} sites coupled by $J'$, 
denote by ${\bf d}$ the vector joining them and label by $\ell$ the $J'$-links they form. 
Note that $J'$-links form a rhombic lattice with coordination number $z=4$. 
With this notation, the tight-binding Bose-Hubbard Hamiltonian \cite{BoseHubbard} with on-site repulsive interactions reads:
\begin{equation}
\label{BHmodel2}
\begin{aligned}
&H =-J'\sum_{\ell} [a_{\ell}^{\dagger} b_{\ell}+b_{\ell}^{\dagger} a_{\ell}]
-J\sum_{\langle\ell,\ell'\rangle}[a_\ell^{\dagger} b_{\ell'}+b_{\ell'}^{\dagger} a_{\ell}]\\
&+\frac{U}{2}\sum_{\ell}[\hat{n}_{a}^{\ell}\left(\hat{n}_{a}^{\ell}-1\right)+\hat{n}_{b}^{\ell}\left(\hat{n}_{b}^{\ell}-1\right)]
-\mu\sum_{\ell}[\hat{n}_{a}^{\ell}+\hat{n}_{b}^{\ell}].
\end{aligned}
\end{equation}
Here $a^\dagger_{\ell}$ ($a_{\ell}$) and $b^\dagger_{\ell}$ ($b_{\ell}$) 
represent the creation (annihilation) operators associated with the endpoint sites \textsc{a} and \textsc{b} of the $J'$-link $\ell$.
The corresponding number operators are $\hat{n}^\ell_{a} = a^\dagger_{\ell} a_{\ell}$ and $\hat{n}^\ell_{b} = b^\dagger_{\ell} b_{\ell}$. 
The (positive) interaction strength is $U$ and $\mu$ is the chemical potential. The summations run over all $J'$-links $\ell$ and, 
in the kinetic term, over their four nearest-neighbor $J'$-links $\ell'$ such that the \textsc{a}-site on $\ell$ and the \textsc{b}-site 
on $\ell'$ are nearest neighbors.

In the following, we investigate the zero-temperature phase diagram of Eq.\eqref{BHmodel2} within a mean-field 
approach~\cite{tcr,ageorge_varenna,ketterle1}. We mainly restrict our analysis to the dimer regime $J'> J$. 
As long as $J'<2J$, the band structure of the non-interacting case ($U=0$) depicts the celebrated conical intersections at the Dirac points 
around $E=0$ and the system is a semi-metal. At $J'=2J$, the two Dirac points merge and  the band structure undergoes a 
topological metal-insulator transition~\cite{Zhu_07,grapheneKL,Montambaux_09}. When $J'>2J$, the band structure 
consists of two bands separated by $2(J'-2J)$. 
When $J'\gg J$, this is simply the energy separation between the symmetric and antisymmetric dimer 
states $|\ell\pm\rangle=\left(|\textsc{a}\rangle_\ell\pm|\textsc{b}\rangle_\ell\right)/\sqrt{2}$ (energy $\mp J'$) built on each $J'$-link $\ell$. 
These dimer states give rise to the two preceding bands, each with a width $4J$ independent of $J'$. In this weak inter-link 
coupling regime (or strong dimer regime), we expect the physics to be driven by the lower band and the MI-SF phase transition to 
be controlled by the ratio $J/U$.
The Mott ground state is then well approximated 
by $\prod_\ell|n, \ell+\rangle$, $|n, \ell+\rangle$ being the Fock state with $n$ bosons in the symmetric state of link $\ell$. 
This state is beyond the reach of the standard Gutzwiller's ansatz which relies on a product of {\it on-site} states. This salient feature directly arises from the 2-point topology of the graphene lattice and cannot happen with the square lattice 
where a strong imbalance of one hopping parameter leads to weakly-coupled 1D chains. We improve Gutzwiller's ansatz by incorporating the correlations between $J'$-link sites and write the ground state as a product of {\it on-link} states $\ket{GS}=\prod_\ell \ket{\ell}$:
\begin{align}
\ket{\ell}&=\sum_{n,m}f_{n,m}^{(\ell)}\ket{n,\textsc{a}; m,\textsc{b}}_{\ell} 
\quad (\mbox{with} \sum_{n,m}|f_{n,m}^{(\ell)}|^2= 1)\label{extGutz} \\
&= \sum_{p,q}g_{p,q}^{(\ell)}\ket{p,+ ; q, -}_{\ell}
\qquad (\mbox{with} \sum_{p,q}|g_{p,q}^{(\ell)}|^2 = 1), \label{extGutzSym}
\end{align}
where
$\ket{n,\textsc{a}; m,\textsc{b}}_{\ell}$ is the Fock state on $J'$-link $\ell$ with $n$ atoms on site \textsc{a} and $m$ atoms on site \textsc{b} while $\ket{p,+ ; q, -}_{\ell}$ is the Fock state on the same $J'$-link $\ell$ with $p$ 
atoms in the symmetric state $\ket{\ell +}$ and $q$ atoms in the anti-symmetric state $\ket{\ell -}$. Relating the $f_{n,m}^{(\ell)}$ and the $g_{p,q}^{(\ell)}$ is easy since the annihilation and creation operators for the $\ket{\ell\pm}$ states are $d^{(\dagger)}_{\ell,\pm}=( a^{(\dagger)}_{\ell}\pm b^{(\dagger)}_{\ell})/\sqrt{2}$.
If both Eqs.\eqref{extGutz}-\eqref{extGutzSym} represent the most general dimer state, and fully describe the system on each $J'$-link, Eq.\eqref{extGutzSym} proves more useful in the limit $J'\gg J$. The ground state of Eq.\eqref{BHmodel2} has been obtained by imaginary-time evolution of an initial state with
random values of amplitudes and periodic boundary conditions~\cite{Sachdev,lewenstein07}. As the structure of the corresponding nonlinear time-dependent equations is rather involved, a $4$th-order Runge-Kutta method was necessary.

\section{Uncoupled dimers}
\label{udim}

\begin{figure*}[thb] 
\includegraphics[width=16cm]{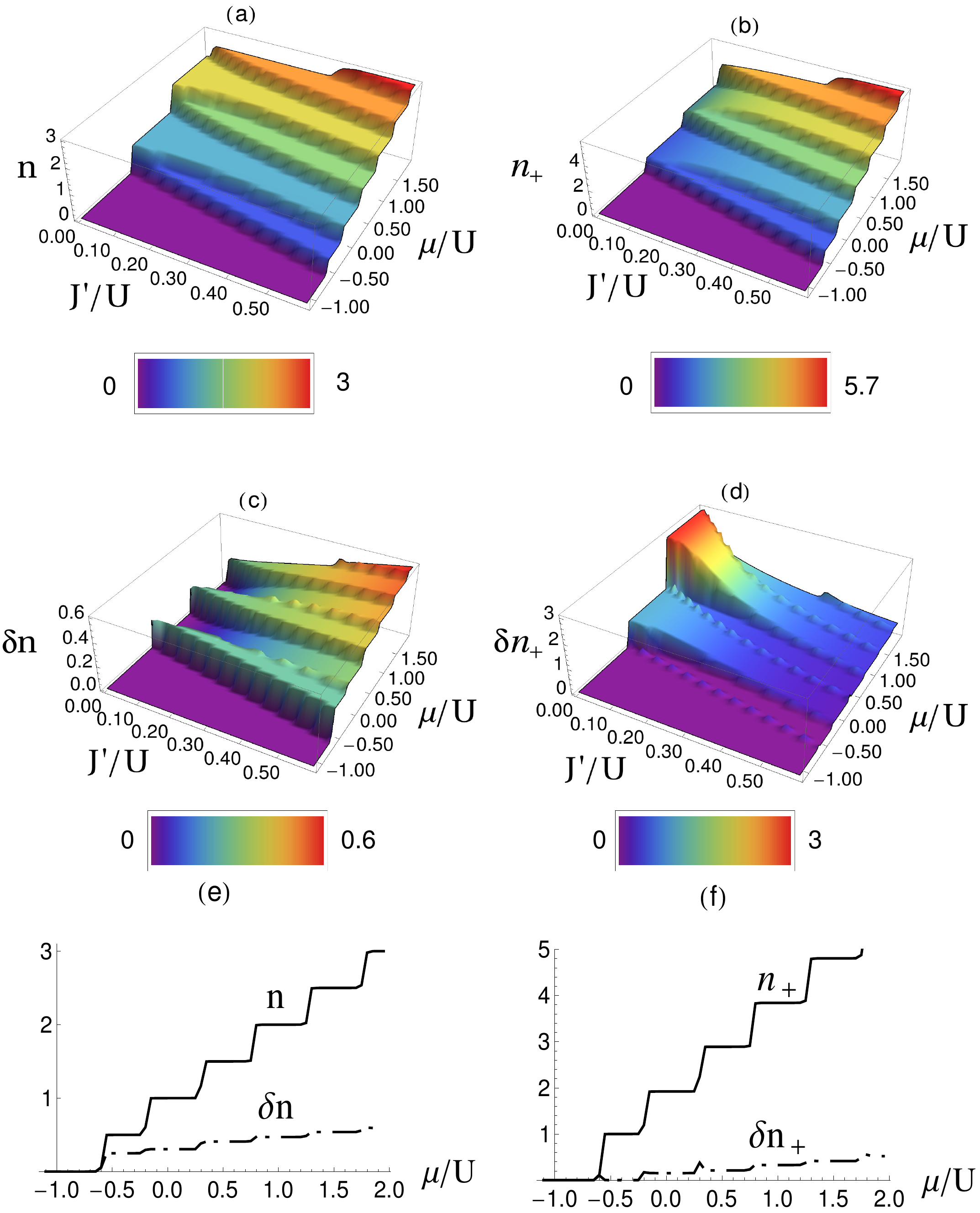}
\caption{(color online). Phase diagram of Hubbard Hamiltonian \eqref{BHmodel2} for $J=0$. 
Left: on-site density $n$ (a) and variance $\delta n$ (c) as a function of $J'/U$ and $\mu/U$ with their color codes. Plot (e) is a cut of (a) and (c) for $J'/U =0.6$. Right: same as left but for the symmetric density $n_+$ and variance $\delta n_+$. For large $J'/U$, 
the on-site density depicts plateaus at half-integer fillings, corresponding to integer occupation numbers of the symmetric states $\ket{\ell+}$. 
The bottom plots emphasize that the first plateau corresponds to the Fock state $\ket{1,+ ; 0,-}_{\ell}$.}
\label{num_diag0}
\end{figure*}

We first look for the ground state of the on-link dimer situation. It corresponds to the Mott states of~Eq.\eqref{BHmodel2}. 
Fig.~\ref{num_diag0} shows the on-site (left) and symmetric (right) average occupation numbers and their variances as  
functions of $\mu$ and $J'$ in units of $U$. 
When $J'/U$ is small, the on-site density depicts the usual Mott plateaus at integer fillings. But, when $J'/U$ increases, 
one clearly observes the appearance of new plateaus at half-integer fillings. At the same time, plateaus at odd integer values $2p+1$ 
start to appear in the symmetric state density. For larger $J'/U$ values, each of these plateaus splits into 
two new ones with $2p+1$ and $2p+2$ fillings. Correspondingly (not shown here), the $\ket{\ell -}$-state occupation number decreases from the same value $2p+1$, at $J'=0$, to an almost vanishing value. The plot of the on-site and symmetric variances 
emphasizes that the ground state of the system evolves from an on-site Fock state to an (almost)
on-link Fock state 
when $J'/U$ increases. Indeed, when $J'\approx 0$, the on-site variance is almost zero while the $\ket{\ell +}$-state variance is the largest. Increasing $J'/U$, the on-site variance gets larger while the symmetric one almost vanishes. More precisely, the ground state is well approximated by Fock states $\ket{n, \textsc{a} ; n, \textsc{b}}_{\ell}$ for $J'\approx 0$, 
whereas it is well approximated by Fock states $\ket{p,+ ; 0, -}_{\ell}$ for larger $J'$.

Note that, for $J'/U=0.6$, the symmetric variance, albeit extremely small, 
is not strictly zero and even increases with $\mu$. The ground state is thus slightly contaminated by the $\ket{\ell -}$ states at finite $J'$ due to the interacting part of the Hamiltonian~\eqref{BHmodel2}. 
Indeed only the latter is not diagonal in the $\ket{\ell\pm}$ basis and couples states $\ket{p,+ ; q,-}_{\ell}$ with the same total number of atoms $p+q$, $p$ and $q$ increments being by steps of two. 
The actual ground state thus reads:
\begin{equation}
\label{GSS}
 \ket{GS}_l=\alpha \ket{p,\!+ ; 0, \!-}_{\ell}+\sum_{q \leq p/2} \alpha_q\ket{p\!-\!2q,\!+ ; 2q,\!-}_{\ell}.
\end{equation}
In the large $J'/U$ limit, $\alpha$ is of the order of unity, 
whereas the other (small) coefficients are becoming smaller with increasing $q$ or $J'/U$. 
Except for $p=0$ and $p=1$, $n^{\ell}_+$ is always a bit less than an integer and the symmetric variance is never strictly zero, even 
if discrepancies go to $0$ when increasing $J'/U$. 
This is clearly seen in Fig.~\ref{num_diag0}f at $J'/U=0.6$ (dot-dashed line) where the variance deviates 
from zero only for plateaus with symmetric filling $p\geq 2$ (solid line). For small $J'/U$ values, 
the ground state simply becomes $\ket{n,\textsc{a} ; n,\textsc{b}}_{\ell}$, such that the coefficients $\alpha_q$ are now given by $(-1)^q/(n-q)!q!$ (up to a normalization factor). They reach a maximum for $q=n/2$, leading to the same 
number of bosons in the $\ket{\ell\pm}$ states. This corresponds to an even-integer 
symmetric filling $p=2n$ in Eq.~\eqref{GSS}. For odd filling $p=2n+1$, 
Eq.~\eqref{GSS}  is not the ground state of the system when $J'=0$. 
However increasing $J'$ lowers the energy of the $\ket{\ell+}$ state, which compensates for the additional energy cost for this extra boson. 
State Eq.~\eqref{GSS} with $2n+1$ bosons then becomes more favorable in a given range of $\mu/U$. 
The actual scenario is of course less simple since, at intermediate values of $J'/U$, 
the ground state does not correspond to a pure Fock state in any one of the two basis. 
For instance, the Mott state at unit filling reads:
\begin{equation}
\begin{aligned}
\label{mott1}
 |M1\rangle&=c|2,+;0,-\rangle-s|0,+;2-\rangle\\
	  &=\frac{c-s}{2}\left(|2,\textsc{a};0,\textsc{b}\rangle+|0,\textsc{a};2,\textsc{b}\rangle\right)+
	  \frac{c+s}{\sqrt{2}}|1,\textsc{a};1,\textsc{b}\rangle\\
	  \text{with  } c^2&=\frac{1}{2}\left[1+\frac{4J'}{\sqrt{(4J')^2+U^2}}\right]\\
	               s^2&=\frac{1}{2}\left[1-\frac{4J'}{\sqrt{(4J')^2+U^2}}\right].
 \end{aligned}
\end{equation}
In the limit $J'\ll U$, one recovers the usual Mott state $|1,\textsc{a};1,\textsc{b}\rangle$, whereas in the limit $J'\gg U$, as explained above,
the Mott state is simply the symmetric state $|2,+;0,-\rangle$. Whereas the onsite density is independent of $J'/U$, the difference between 
the symmetric and antisymmetric occupation numbers is $2(c^2-s^2)=8J'/\sqrt{(4J')^2+U^2}$, ranging from $0$ for $J'\ll U$ to $2$ for $J'\gg U$.
As explained, in Sec.\ref{sec:expsign}, this feature can be inferred from the velocity distribution of the atoms.

\section{Coupled dimers}
\label{cdim}

\begin{figure*}[thb]
\includegraphics[width=16cm]{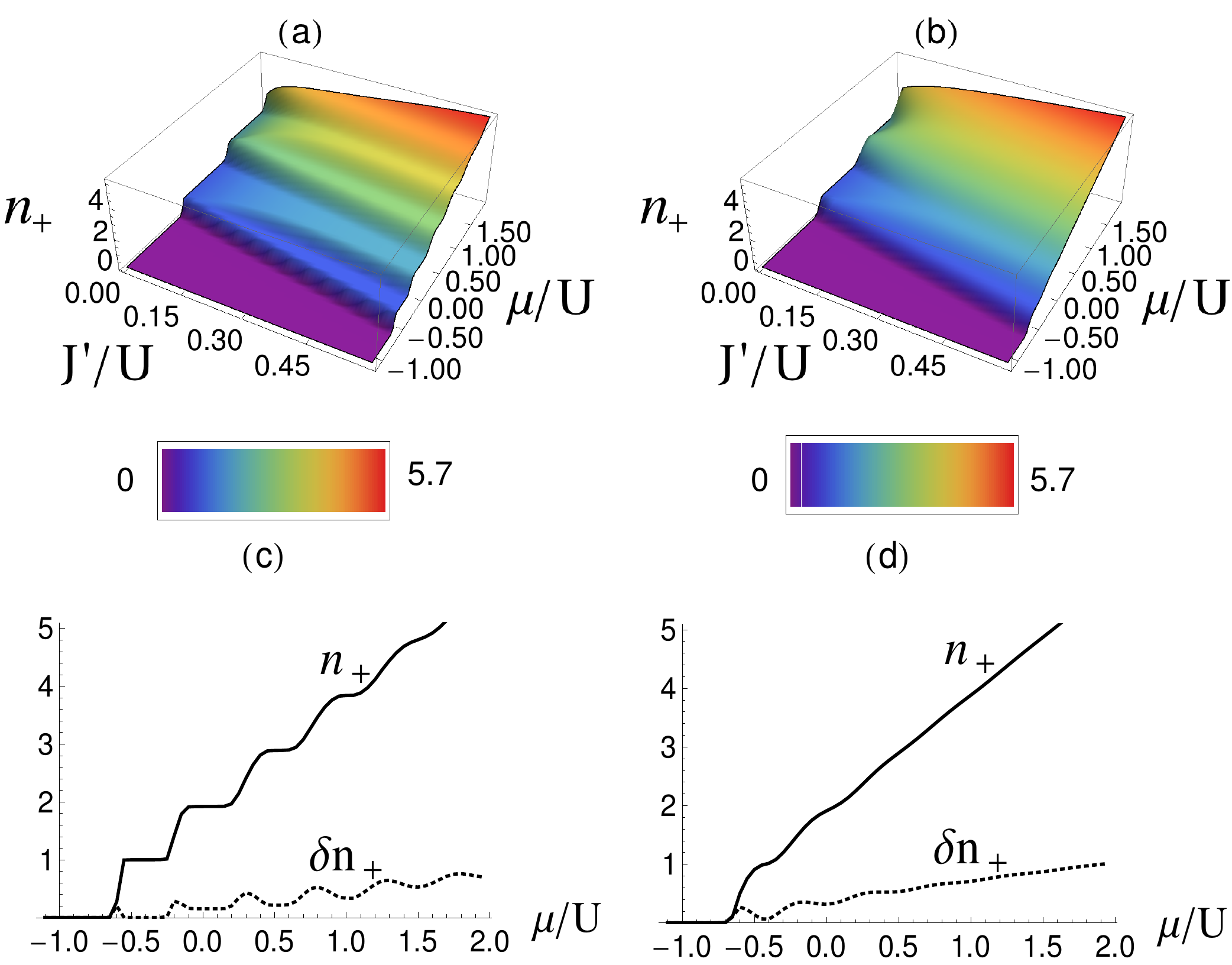}
\caption{(color online). Phase diagram of Hubbard Hamiltonian \eqref{BHmodel2}. Left column: $J/U=0.015$. Right column: $J/U=0.04$. (a), (b): occupation number $n_+$ as a function of $J'/U$ and $\mu /U$ with the color code. (c), (d): $n_+$ and $\delta n_+$ as a function of $\mu/U$ for $J'/U=0.6$. For $J/U=0.015$, the first three Mott plateaus are clearly visible whereas 
they are smoothed out for $J/U=0.04$. }
\label{num_diag002}
\end{figure*}

We now consider the $(J'/U,\mu/U)$ 
phase diagram obtained in the $\ket{\ell +}$ states when $J$ is non zero. As seen in Fig.~\ref{num_diag002}a and Fig.~\ref{num_diag002}b, the situation is very much similar to the usual 
MI-SF transition. One finds regions with constant symmetric occupation numbers $n_+$ and almost vanishing variance, corresponding 
to a Mott state with vanishing compressibility $\chi = \partial n/\partial \mu$.
These Mott lobes are surrounded by a superfluid 
sea where $\chi$ is finite. For small $J/U$, the lower Mott plateaus are still well visible, whereas the 
higher plateaus are almost all smoothed out; see Fig.~\ref{num_diag002}c 
where the variance of the first two Mott phases is still small. Between the plateaus, the variance has maxima revealing the 
superfluid phases. For $\mu\agt U$, $n_+$ varies smoothly and the system is superfluid. 
For larger value of $J/U$, the half-integer Mott lobes almost entirely disappear except at very low values of $J'/U$, 
see Fig.~\ref{num_diag002}b.
This is further exemplified by Fig.~\ref{num_diag002}d
where both $n_+$ and $\delta n_+$ vary smoothly as $\mu$ is increased. The evolution of the $g^{(\ell)}_{p,q}$ amplitudes when $J$ increases at fixed $J'$ is similar to the MI-SF scenario in the usual Gutzwiller's ansatz. When $J'/U$ is large, $g^{(\ell)}_{p,q} = \delta_{pp_0}\delta_{q0}$ and the Mott state is approximated by the Fock state $\ket{p_0,+ ; 0,-}_{\ell}$. At the transition, the distribution $g^{(\ell)}_{p,q}$
starts broadening, but only in the $\textit{symmetric}$ direction $q=0$ and the physics takes  place entirely in the 
symmetric subspace. For instance, when $J'\gg J\gg U$, the superfluid phase at density $\rho$ is described by the coherent state $\ket{\sqrt{\rho}}_+\otimes\ket{0}_-$ with no bosons 
in the anti-symmetric subspace.  For intermediate $J'/U$, the Mott state 
is slightly contaminated by anti-symmetric contributions. The MI-SF transition scenario 
remains however the same: the $g^{(\ell)}_{p,q}$ still spread along 
the symmetric direction and keep a structure similar to Eq.~\eqref{GSS}.

\begin{figure}[thb]
 \includegraphics[width=8cm]{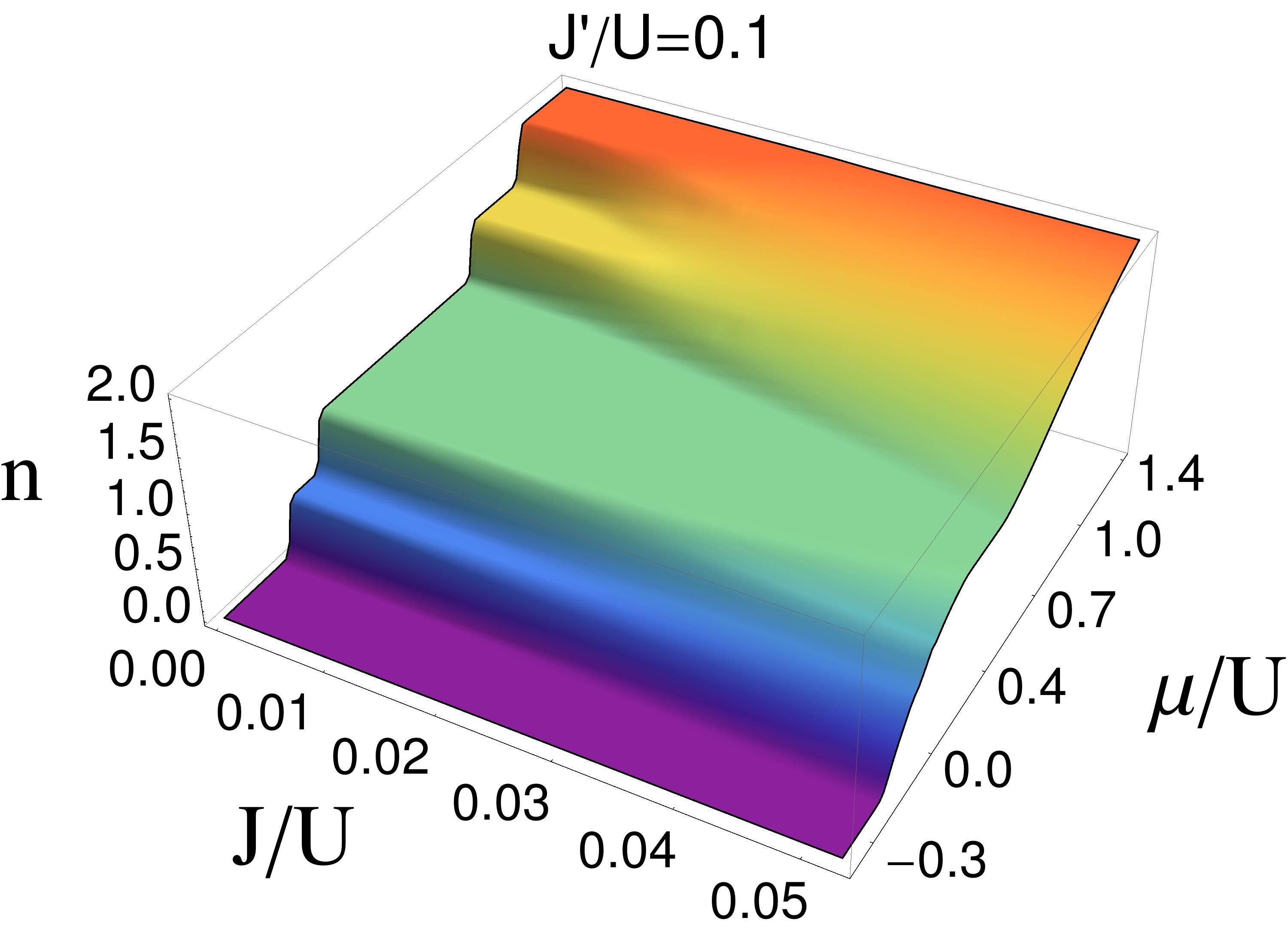}
 \caption{(color online). Phase diagram in the $(J/U,\mu/U)$ plane for a fixed value of $J'/U=0.1$. For this small value of $J'$, 
 the dominant Mott phases correspond to integer fillings, whereas the half-integer Mott phase depicts a much smaller extension. For instance,
 the $n=0.5$ Mott lobe corresponds to $-J'<\mu<J'$, for $J=0$, whereas the tip of the lobe corresponds to $J\approx J'/4$. \label{Jp0_1}}
\end{figure}

\begin{figure}[thb]
 \includegraphics[width=8cm]{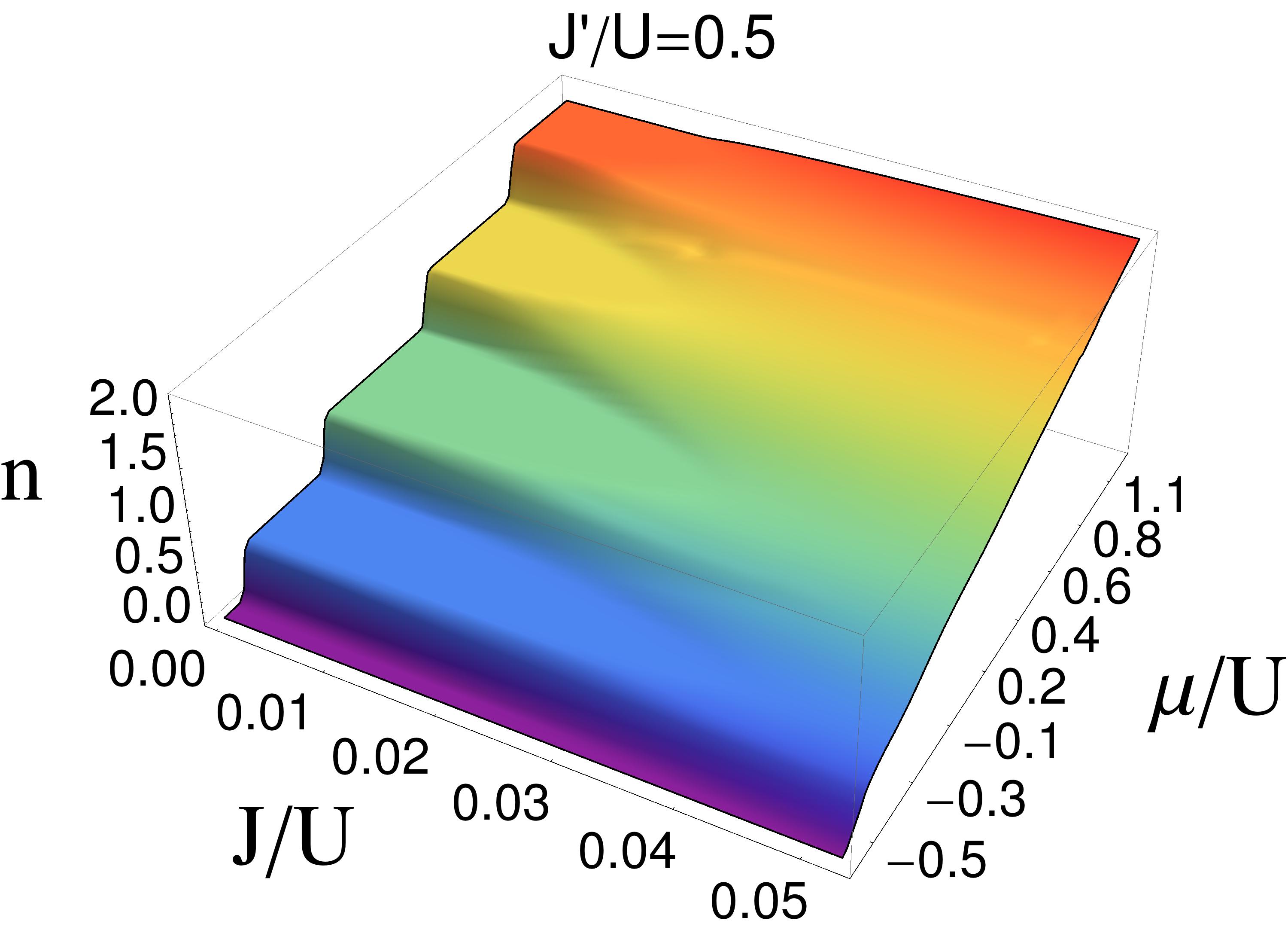}
 \caption{(color online). Phase diagram in the $(J/U,\mu/U)$ plane for a fixed value of $J'/U=0.5$. The half-integer Mott
 phases have a larger extension, but still smaller than the integer ones.
 \label{Jp0_5}}
\end{figure}

\begin{figure}[thb]
 \includegraphics[width=8cm]{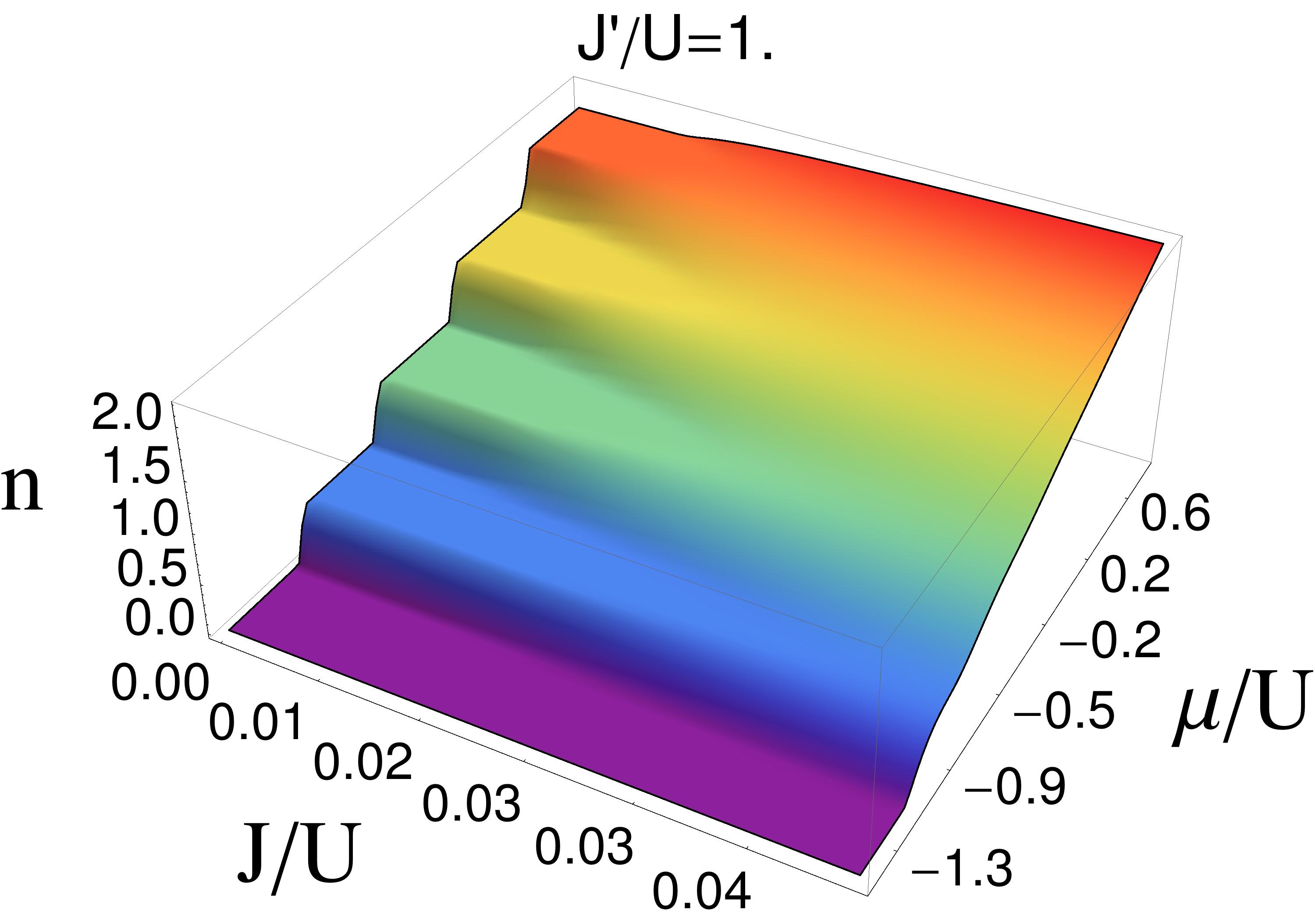}
 \caption{(color online). Phase diagram in the $(J/U,\mu/U)$ plane for a fixed value of $J'/U=1$. As explained in the text, the situation
 corresponds to weakly coupled dimers, such that the phase diagram resembles the one of the square lattice: all Mott lobes have the
 same width $U$ (for $J=0$), the transition from the Mott phase with $p$ bosons in the symmetric state, corresponding to a lattice filling $p/2$,
to the Mott phases with $p+1$ bosons occurs for $\mu=-J'+pU/2$. The tip of the Mott lobe with $p$ bosons in the symmetric state
is given by $U/J^{(p)}_{c4} = 4(2p+1 +2\sqrt{p(p+1)})$.
\label{Jp1_0}}
\end{figure}

The phase diagrams in the $(J/U,\mu/U)$ plane are shown in Fig.~\ref{Jp0_1},  Fig.~\ref{Jp0_5} and Fig.~\ref{Jp1_0}, for three
different values of $J'$, namely $J'/U=0.1$, $J'/U=0.5$ and $J'/U=1$. As expected, the size of the half-integer Mott plateau increases 
with increasing $J'$, whereas the size of the integer ones is decreasing. 
More precisely, in the large $J'/U$ limit, the different Mott states are simply $|p,+;0-\rangle$, with a free energy given by 
$E_p^+=Up(p-1)/4-(J'+\mu) p$. Therefore, along the $J=0$ axis, the transition between the $p$ and $p+1$ Mott phases occurs for 
$\mu=-J'+pU/2$, such that all Mott states have the same width $U/2$. For an arbitrary value of $J'$, the transition, along the
$J=0$ axis, between the different Mott phases is more complicated to determine, due to the non-trivial structure of the Mott states,
see Eq.~\eqref{GSS}. Nevertheless, in the case of the half-integer Mott phase, one can simply obtain the $J=0$ boundary:
$-J'<\mu<J'+U/2-\sqrt{U^2+(4J')^2}/2$. In particular, in the limit $J'\ll U$, one obtains: $-J'<\mu<J'$, in agreement with 
Fig.~\ref{Jp0_1}. In addition, one can see that the half-integer Mott lobe always corresponds to $J$ values   much lower than $J'$, i.e. the 
situation of weakly coupled dimers.

The preceding analysis shows that the half-integer Mott lobes are observed in the regime
$J\ll J'\lesssim U$ and that, for a fixed value of $J'$, the Mott-superfluid transition occurs for values of the parameter 
$J/U\approx 10^{-2}$, similar to the usual Mott-superfluid transition. In addition, deep inside the Mott phase, 
the gap is expected to be of the order of $J'$ or larger. Therefore, for temperatures such that $k_BT\ll J$,
its impact on the system properties is expected to be comparable to the standard situation, and thus, the half-integer Mott lobes
are within experimental reach. A full description of these 
finite temperature effects can be obtained from the mean-field excitations spectrum, i.e. 
the Bogoliubov modes, but this is beyond the scope of the paper.

\section{Critical hopping amplitude}
\label{crithop}

\begin{figure}[thb] 
\includegraphics[width=8cm]{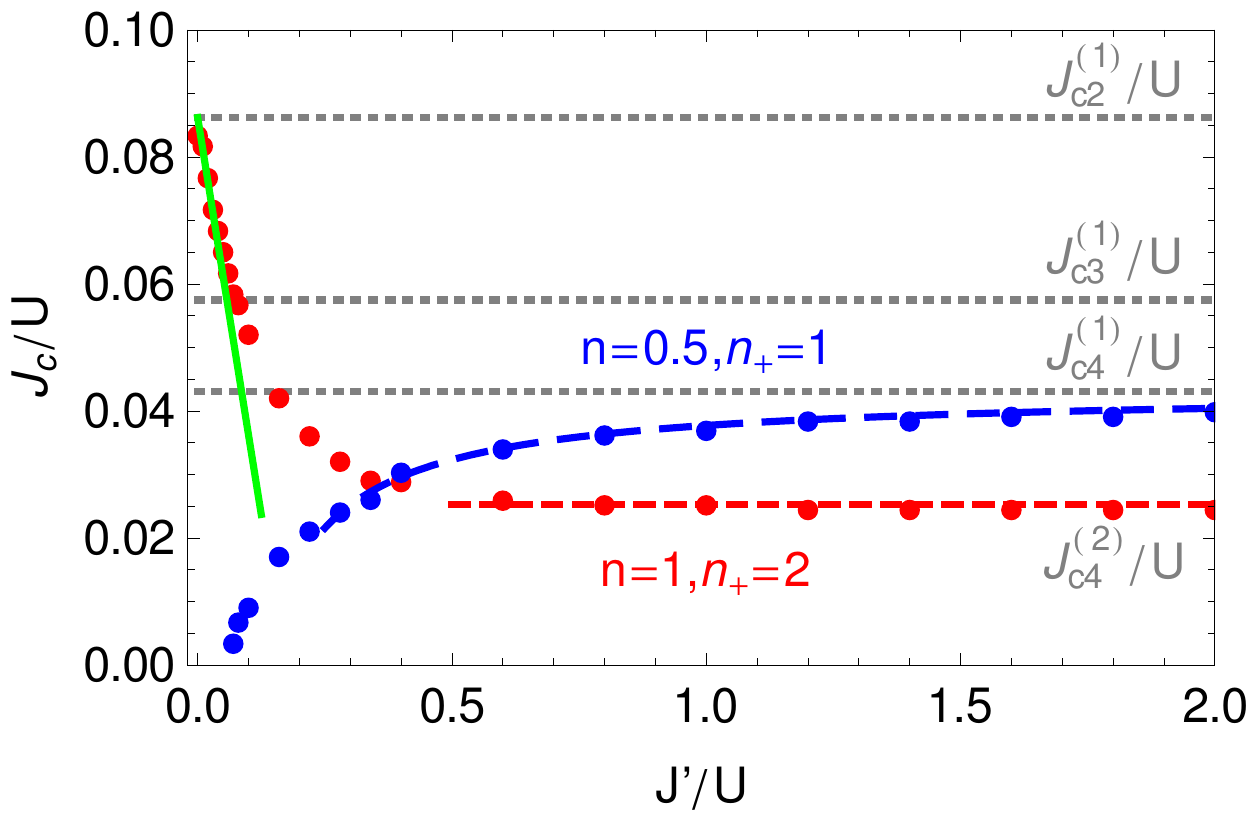}
\caption{(color online) Critical hopping $J_c$ for the MI-SF transition as a function of $J'$ (in units of $U$). Red (resp. blue) dots are numerical data obtained with the extended Gutzwiller's ansatz for the $n=1$ (resp. $n=0.5$) Mott lobe. Continuous 
and long-dashed lines are analytical predictions. The horizontal black dashed lines correspond to the critical values 
$J^{(1)}_{cz}$ ($z=1,2,3$). See text for details.}
\label{Jcr}
\end{figure}

We obtain the MI-SF critical hopping rate $J_{c}$
by monitoring the 
tip of the two first Mott lobes, see Fig.~\ref{Jcr}. For integer filling,  the usual Bose-Hubbard model predicts 
$U/J^{(\rho)}_{cz} = z(2\rho+1 +2\sqrt{\rho(\rho+1)})$ where the average density $\rho$ is here an integer and $z$ is the lattice coordination number~\cite{tcr,ageorge_varenna}. For small $J'/U$, we have $z=2$ (almost independent 1D chains) 
and the physics is driven by on-site Fock states, the main Mott lobe being at $\rho=n=1$. 
For imbalanced hopping parameters, one gets $(2J+J')=U/5.8$ and thus $J_c= J^{(1)}_{c2} - J'/2$. This prediction 
correctly reproduces our numerical results at small $J'/U$. In the large $J'/J$ limit however, the 
physics takes place in the symmetric subspace and $z=4$ (dimer lattice). It is easy to see from Eq.~\eqref{BHmodel2} that
the effective hopping parameter in the dimer lattice is $J/2$ giving rise to a non-interacting band 
with finite width $4J$ independent of $J'$, such that in the large $J'/U$ limit, $J_c$ for the Mott lobe at  $\rho =n_+=1$ (equivalently $n=0.5$)
reaches the value $J^{(1)}_{c4}$. In addition, for large $J'/U$, 
$J_c$ for the Mott lobe at $n=1$ ($n_+=2$) saturates, as it should, at $J^{(2)}_{c4}$ ($z=4$, $\rho =n_+=2$). 

As explained above, the half-integer Mott state is simply $|1,+;0-\rangle$, such that the boundary between the Mott and 
the superfluid phase can be obtained analytically. More precisely, close to the boundary, a first order perturbation theory leads to the
following expression for the ground state (on a given link $\ell$):
\begin{equation}
\label{fop}
 |\psi\rangle=|1,+;0-\rangle+\sum_k |k\rangle\frac{\langle k|V|1,+;0-\rangle}{E_{10}-E_k},
\end{equation}
where $|k\rangle$ are the different states coupled to $|1,+;0-\rangle$ by the mean-field kinetic energy term 
\begin{equation}
 V=-2J \left(\langle d_{+}\rangle d_{+}^{\dagger}+\langle d_{+}^{\dagger}\rangle d_{+}-
 \langle d_{-}\rangle d_{-}^{\dagger}-\langle d_{-}^{\dagger}\rangle d_{-}\right).
\end{equation}
The states coupled by $V$ are therefore $|1,+;1-\rangle$, $|2,+;0-\rangle$  and $|0,+;0-\rangle$, but the states $|k\rangle$ in Eq.~\eqref{fop}
must be eigenstates of the on-link Hamiltonian, such that the relevant states are:
\begin{equation}
\begin{aligned}
 &|0,+;0-\rangle\\
 &|1,+;1-\rangle\\
 &|+\rangle=c |2,+;0-\rangle-s |0,+;2-\rangle\\
 &|-\rangle=s |2,+;0-\rangle+c |0,+;2-\rangle,
 \end{aligned}
\end{equation}
where $c$ and $s$ are defined in Eq.~\eqref{mott1}. $|+\rangle$ is nothing but the $n=1$  Mott state, an eigenstate of the on-link Hamiltonian. 
The notation $\langle d_{\pm}\rangle$ corresponds to the ground state average value  $\lambda_{\pm}=\langle \psi | d_{\pm}|\psi \rangle$, which must
be self-consistently obtained using the ground state expression given by Eq.~\eqref{fop}~\cite{Jaksch98,ageorge_varenna}. 
$\lambda_{\pm}$ are precisely the mean-field order parameters, with vanishing values 
in the Mott state, and non-zero values in the superfluid phase. From the first order perturbation expression~\eqref{fop}, one obtains that
the boundary between the two phases is given by the following equation:
\begin{equation}
\label{Mhb}
 \frac{1}{4J}=\frac{1}{2}\frac{1}{J'+\mu}+\frac{c^2}{J'-\mu+E_-}+\frac{s^2}{J'-\mu+E_+},
\end{equation}
where $E_{\pm}$ are the energies of the states $|\pm\rangle$, namely $E_{\pm}=\frac{U}2\pm\frac{1}{2}\sqrt{U^2+(4J')^2}$. 
For $J\rightarrow0$, the boundary corresponds to either $\mu=-J'$ or $\mu=J'+E-$,  which are exactly the two values given in the preceding section 
(since $E_+>E_-$, the boundary is given by the $E_-$ term). 
The critical value of $J_c$, i.e. the tip of the Mott lobe, is the largest $J$ value on the boundary and is thus given by the minimum value of the 
right-hand side of Eq.~\ref{Mhb}, for 
$-J'<\mu<J'+E_-$.   
In the small $J'/U$ limit, the critical value $J_c$ for the half-integer
Mott phase scales like $J'/4$, a value within experimental reach, see Ref.~\cite{Esslinger3} where $J'/J=10$ has been achieved.
In the large $J'/U$ limit, one obtains that $J_c = J^{(1)}_{c4}\left(1-\frac{1}{8}\frac{U}{J'}\right)$, 
in good agreement with our numerical data for $\rho =n_+=1$ (equivalently $n=0.5$), see the blue dashed line in Fig.~\ref{Jcr}.

In principle, the boundary between the superfluid and the Mott phase for higher fillings could be obtained in a similar way, the final expressions are
quite involved, and therefore not put in this paper.

Finally, 
even though the present ansatz favors $J'$-links over $J$-links, it is quite remarkable that our numerical results for the Mott lobe at $\rho=n=1$ cross the 
critical value $J^{(1)}_{c3}$ of the balanced honeycomb lattice roughly when $J_{c}=J'$. From that point of 
view, the system undergoes a cross-over from a quasi-1D situation with two neighbors (weakly-coupled chains) to a 2D situation 
with four neighbors, the balanced honeycomb lattice with 3 neighbors being the intermediate situation. 

\section{Experimental signatures}
\label{sec:expsign}

\begin{figure}[thb]
\includegraphics[width=7cm]{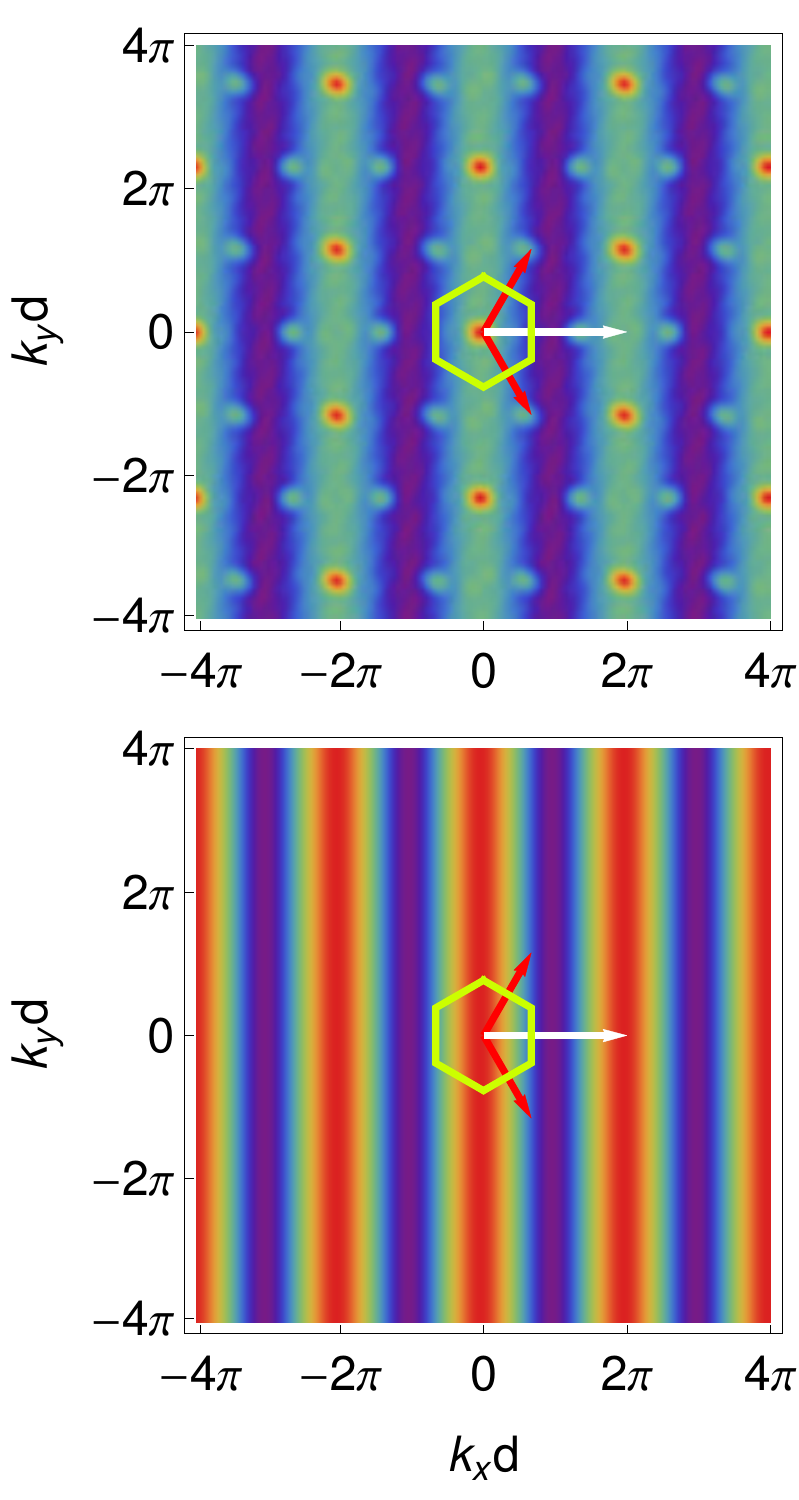}
\caption{(color online) Momentum distribution $n_{\mathbf{k}}$ for $J/U=0.015$, $J'/U=0.6$. Top frame: SF phase ($\mu/U=-0.14$). 
Bottom frame: MI phase ($\mu/U=-0.5$). The hexagon is the first Brillouin zone with its reciprocal lattice vectors 
(red arrows). The $J'$-link ${\bf d}$ vector is chosen along $Ox$. The periodic stripes (with period $2\pi/d$ shown by the white arrow) 
are a signature of the quantum correlations between the \textsc{a} and \textsc{b} sites along $J'$-links. See text for details.
}
\label{num_mom}
\end{figure}

The momentum distribution of the atoms, measured after the optical 
lattice is rapidly switched off, is known to exhibit a clear signature of the MI-SF transition~\cite{ReviewBloch,expBloch}.
One can show that
\begin{equation}
 n_{\mathbf{k}} \propto \sum_{i,j} \langle GS|c_i^{\dagger}c_j|GS\rangle \exp{\left[i\mathbf{k}\cdot(\mathbf{R}_j-\mathbf{R}_i)\right]},
\end{equation}
where $\mathbf{R}_i$ is the position of site $i$. Since the usual Guztwiller's ansatz discards inter-site correlations, only terms 
like $\langle c_i^{\dagger}c_i\rangle $ or $\langle c_i^{\dagger}\rangle\langle c_j\rangle \exp\left[i\mathbf{k}\cdot(\mathbf{R}_j-\mathbf{R}_i)\right]$ 
contribute to $n_{\mathbf{k}}$. Our extended ansatz includes the additional terms 
$\langle a_{\ell}^{\dagger}b_{\ell}\rangle\exp(i\mathbf{k}\cdot\mathbf{d})$
which give rise to a periodic modulation of the velocity distribution, a smoking-gun of a Mott phase built on a 
symmetric state. For a pure Fock state $\ket{p,+; 0,-}$, $n_{MI}(\mathbf{k})=p\left(1+\cos{(\mathbf{k}\cdot\mathbf{d})}\right)$ and the modulation 
contrast is $C=1$. 
With the actual ground state of Eq.~\eqref{GSS}, $C\leq1$. In general, measuring the contrast $C$ for Mott states with $n\geq 1$ would reveal that 
they are not simple Fock states with fixed on-site density but have an underlying structure. For instance, $C=J'/\sqrt{J'^2+(U/4)^2}$ for 
the Mott state $n=1$.
For the superfluid state, a product of on-site coherent states, one finds $n_{SF}(\mathbf{k})=\rho_0\sum_{ij}e^{i\mathbf{k}\cdot(\mathbf{R}_j-\mathbf{R}_i)}$ 
(assuming a uniform density $\rho_0$), which depicts peaks at the reciprocal lattice vectors modulated by the square of the structure factor of the lattice. 
The additional on-link correlations thus show up as an additional modulation on top of this ideal distribution. 
All these properties are confirmed by our numerical calculations, see Fig.~\ref{num_mom}, where 
both the superfluid (top plot) and the Mott (bottom plot) phases display a periodic modulation along ${\bf d}$ with period $2\pi/d$ ($d=|{\bf d}|$).
Note that, besides the preceding modulation, the effect of the structure factor 
in the superfluid phase is clearly visible in the different peak heights.

\section{Conclusion}
\label{conc}

In conclusion, using an extended Gutzwiller's ansatz, we have described the  properties of the MI-SF transition of ultracold bosons in a honeycomb lattice.
We have found Mott phases at \textit{half-integer} fillings, arising directly from the interplay between quantum correlations and the topology of 
the honeycomb lattice. Future work will address the excitations of the system~\cite{exp1D} as they can lead to additional experimental signatures. 
Finally, it would be interesting to study the impact of an external (non-Abelian) gauge field on the properties 
of the ground state~\cite{Gorecka_11,dalibard1,Bermudez10}.

The Centre for Quantum Technologies is a Research Centre of 
Excellence funded by the Ministry of Education and National Research Foundation of Singapore.  ChM is a Fellow of the Institute of Advanced Studies~(NTU).

\end{document}